\documentclass[a4paper]{article}

\RequirePackage[OT1]{fontenc}
\RequirePackage{amsthm,amsmath}
\RequirePackage[numbers]{natbib}
\RequirePackage[colorlinks,citecolor=blue,urlcolor=blue]{hyperref}
\usepackage{latexsym,graphicx,subcaption,verbatim,nameref}
\usepackage{diagbox}
\usepackage[all,cmtip]{xy}
\usepackage[toc,title,titletoc]{appendix}
\usepackage[capposition=top]{floatrow}
\usepackage{rotating}
\usepackage{threeparttable,booktabs}
\usepackage{csquotes} 
\usepackage{chngcntr}
\usepackage{subcaption}
\usepackage[affil-it]{authblk}
\usepackage{graphicx}
\usepackage{esint}
\usepackage{relsize}
\usepackage{color}
\usepackage{algorithm}
\usepackage{algpseudocode}
\counterwithin{figure}{section}
\counterwithin{table}{section}
\numberwithin{equation}{section}
\theoremstyle{plain}
\newtheorem{theorem}{Theorem}[section]

\newtheorem{corollary}[theorem]{Corollary}

\theoremstyle{definition}

\newcommand*{\Scale}[2][4]{\scalebox{#1}{$#2$}}%
\renewcommand{\footnotesize}{\fontsize{8.5pt}{10pt}\selectfont}
\begin{document}

%\begin{frontmatter}
\title{Insert \enquote{Price} to Coxian Phase-Type Models: An Application to Hospital Charge and Length of Stay Data}%\thanksref{T1}}
%\runtitle{Stochastic Process Model and Health Data}
%\thankstext{T1}{}%Footnote to the title with the ``thankstext'' command.}

%\begin{aug}

\author{Yanqiao Zheng%
\thanks{Email: \texttt{yanqiaoz@buffal.edu}}}
\affil{School of Finance,\\ Zhejiang University of Finance and Economics}
\affil{Department of Economics,\\ University at Buffalo, SUNY}

\author{Xiaoqi Zhang%
\thanks{Email: \texttt{xiaoqizh@buffal.edu}; Corresponding Author.}}
\affil{School of Finance,\\ Zhejiang University of Finance and Economics}
\affil{Department of Mathematics,\\ University at Buffalo, SUNY}
\maketitle
%\and
%\author{\fnms{John} \snm{Ringland}\thanksref{b,e2}\ead[label=e2,mark]{ringland@buffalo.edu}}
%
%
%\address[a]{Math Building 244\\
%State University of New York at Buffalo\\
%Buffalo, NY, 14260\\
%\printead{e1}}
%\address[b]{Math Building 244\\
%State University of New York at Buffalo\\
%Buffalo, NY, 14260\\
%\printead{e2}}
%
%
%\runauthor{X. Zhang and J. Ringland}
%
%\affiliation{Math Department, State University of New York at Buffalo}
%
%%
%%\author{\fnms{Xiaoqi} \snm{Zhang}\thanksref{m1}\ead[label=e1]{xiaoqizh@buffalo.edu}}
%%%{t1,m1}\ead[label=e1]{xiaoqizh@buffalo.edu}
%%%\ead[label=u1,url]{http://www.foo.com}}
%%\and
%%\author{\fnms{John} \snm{Ringland}\thanksref{m2}\ead[label=e2]
%%{ringland@buffalo.edu}}
%%
%%
%%%\thankstext{t1}{}
%%
%%\runauthor{X. Zhang and J. Ringland}
%%
%%\affiliation{Math Department, State University of New York at Buffalo \thanksmark{m1}}
%%\affiliation{Math Department, State University of New York at Buffalo\thanksmark{m2}}
%%
%%\address{Math Building 244\\
%%State University of New York at Buffalo\\
%%Buffalo, NY, 14260\\
%%\printead{e1}%\printead*{e2}\\
%%\phantom{E-mail:\ }}
%
%%\address{Address of the Third author\\
%%Usually a few lines long\\
%%Usually a few lines long\\
%%\printead{e3}\\
%%\printead{u1}}
%\end{aug}

\begin{abstract}
In this paper, we discuss the connection between the RGRST models \cite{gardiner2002longitudinal,polverejan2003estimating
,xqzhang2017} and the Coxian Phase-Type (CPH) models \cite{Tang2012thesis,marshall2007estimating} through a construction that converts a special sub-class of RGRST models to CPH models. Both of the two models are widely used to characterize the distribution of hospital charge and length of stay (LOS), but the lack of connections between them makes the two models rarely used together. We claim that our construction can make up this gap and make it possible to take advantage of the two different models simultaneously. As a consequence, we derive a measure of the \enquote{price} of staying in each medical stage (identified with phases of a CPH model), which can't be approached without considering the RGRST and CPH models together. A two-stage algorithm is provided to generate consistent estimation of model parameters. Applying the algorithm to a sample drawn from the New York State's Statewide Planning and Research Cooperative System 2013 (SPARCS 2013), we estimate the prices in a four-phase CPH model and discuss the implications.
\end{abstract}

%\begin{keywords}
%Coxian Phase-Type model, stochastic process, hospital charge, length of stay, price of medical stages
%\end{keywords}

\section{Introduction}
In recent years, the continuous time Phase-Type Markov chain (PH) model has
become popular in the study of hospital charge and length of stay (LOS) data.
Many authors focus in particular on a special sub-class of PH model/distribution, namely
the Coxian Phase-Type (CPH) model/distribution \cite{Tang2012thesis, faddy2009modeling,marshall2007estimating,
marshall2005length,marshall2002modelling,
fackrell2009modelling}  which includes the Erlang
and Gamma distributions as two important subclasses. Unlike other popular theoretical distributions widely used in inpatient
data, such as log-normal and gamma distribution, the CPH model/distribution provides not only a theoretical distribution that can be used
to fit the empirical data, but also give us a sketch of the treatment dynamics that patient experience in hospital. In fact, from CPH models, we can track the pathways that patient went through different medical stages (characterized by the discrete
set of states in the underlying Markov chain) during a hospital
stay. Those pathway information make it possible to clustering patients and facilitate the use of healthcare process improvement technologies, such as Lean Thinking or Six Sigma \cite{mcclean2005markov}.

The other popular approach to study hospital charge and LOS is through the Random Growth with Random Stopping Time (RGRST) model \cite{gardiner2002longitudinal,gardiner2006dynamic, polverejan2003estimating}, which is a class of continuous-time and continuous-state-space stochastic process models. The formal definition of the RGRST models is given as below:

\begin{equation} \label{defn1}
Y_{t}\left(\omega\right)=Y_{0}\left(\omega\right)+
\int_{0}^{t}I\left(\omega,Y_{s},s\right)
\epsilon_{s}\left(\omega\right)ds
\end{equation}
where the process $\{Y_t:\,t\in[0,\infty)\}$ represents the actual charge level at each time. $\left\{ \epsilon_{t}\right\} $
is a non-negative process characterizing the potential increment rate of
charge per unit time provided that patient decides to stay.  
$I\left(.,Y_{s},s\right)$ (taking value in $\left\{ 0,1\right\}$)
is the decision process representing patient's discharge decision, whether
or not to stay in hospital for longer time at each time point $s$, it
takes value 1 if patients decide to stay and 0 otherwise. The decision process $I$ is required to be non-increasing in the following sense:

\begin{equation}\label{non-increasing property}
s\leq s'\Longrightarrow I\left(\omega,Y_{s}\left(\omega\right),s\right)\geq I\left(\omega,Y_{s'}\left(\omega\right),s'\right)\; a.s.
\end{equation}
Like CPH models, the RGRST models do also capture the treatment dynamics that patient experience in hospital. But in contrast to tracking the pathways of patient moving through different medical stages, the RGRST models focus more on describing how patient and/or doctor makes the discharge decision in react to the change of actual charge level and the length of time that patient has stayed in hospital (reflected on the decision process $I$). Therefore, the story of RGRST models is more about the behavioral patterns of patient/doctor behind the treatment dynamics, and the story of CPH models is more on the medical side.

It is natural to think of the possibility
to combine CPH models and RGRST models together in order to extract more information regarding the discharge decision-making on different medical stages. This paper shows that there is a natural way to convert a special sub-class of RGRST models to CPH models, and presents an algorithm to estimate the transition matrix of the CPH model converted from a given RGRST model and the parameters involved in that RGRST model. It turns out that the correspondence between RGRST models and CPH models we build in this paper could provide a measure of the \enquote{price} that patients would like to pay to stay in each medical stage at each time. This price information might be important for the purpose of insurance payment and healthcare process improvement.

The organization of the paper is that: In section 2, we present the correspondence between RGRST models and CPH models and briefly introduce the estimation algorithm. In section 3, we discuss the choice of the parametric form of relevant functions and fit our model to a sample drawn from the database, New York State's Statewide Planning and Research Cooperative System 2013 (SPARCS 2013). We will plot the estimated \enquote{prices} of each medical stage identified as phases in the estimated CPH model and discuss their implications. Section 4 concludes.
\section{Connection to Coxian Phase-Type model}

\subsection{Correspondence between RGRST and CPH}
We need the following two important conditional expectations for the proof of Theorem \ref{connection to CPH}:

\begin{equation} \label{unique eq}
\begin{aligned}
\tilde{q}\left(y,t\right):=&
E\left(\epsilon_{t}\mid G_t=y\right)\\
\tilde{\rho}\left(y,t\right):=&
E\left(I\left(.,Y_{t},t\right)\mid G_t=y\right)
\end{aligned}
\end{equation}
where $G_t
:=Y_0+\int_{0}^{t}\epsilon_{s}ds$ is defined to be the potential growth process of charge. In addition, we denote $\tilde{p}\left(.,t\right)$ as the probability density with respect to $G_t$. Under these notations, the joint PDF of charge and LOS can be expressed as below:

\begin{equation} \label{joint density 2nd expression}
P\left(y,t\right)=\tilde{p}\left(y,t\right)\cdot\left(-\frac{\partial\tilde{q}_{1}}{\partial y}\cdot\tilde{q}-\frac{\partial\tilde{q}_{1}}{\partial t}\right)\left(y,t\right)
\end{equation}
The detailed derivation of Equation \ref{joint density 2nd expression} can be found in \cite{xqzhang2017}.

The main result of this section is that there does exist a correspondence between CPH and RGRST models. The correspondence is built through converting
the continuous variable, charge, in a RGRST model to finite many discrete states
such that the resulting evolution of the probability mass over those discrete states is exactly determined by
the desired CPH model. More precisely, we have the following theorem:
\begin{theorem} \label{connection to CPH}
Fix a RGRST process $\left\{ Y_{t}\right\} $ represented as a triple
$\left(p\left(.,0\right),\tilde{q},\tilde{\rho}\right)$. Suppose
functions $\tilde{q}$, $\tilde{\rho}$ and $p\left(.,0\right)$ are smooth and $\tilde{q}$, $\tilde{\rho}$ satisfy:

\begin{equation}\label{condition 2.6.9}
\begin{aligned}
\tilde{\rho} > & 0\\
\frac{\partial ln\left(\tilde{\rho}\right)}{\partial y}\cdot\tilde{q}+\frac{\partial ln\left(\tilde{\rho}\right)}{\partial t} \equiv & -c
\end{aligned} 
\end{equation}
for some constant $c>0$. Then for any fixed positive integer $n$,
an $n$-dimensional vector $\alpha>0$ with $\sum_{i=1}^{n}\alpha_{i}=1$
and an $n-1$-dim vector $\lambda>0$, there exists an $n-$partition
of the space $[0,\infty)^{2}$ denoted as $\mathcal{P}$ such that
the following time dependent probability mass function $P\left(t\right)$
defined on the $n+1$ tuple $\left\{ 1,\dots,n+1\right\} $ 

\begin{eqnarray*}
P_{i}\left(t\right) & := & Prob\left(Y_{t}\in\mathcal{P}_{i}\cap[0,\infty)\times\left\{ t\right\} ,I\left(.,Y_{t},t\right)=1\right), \text{ } i\in\left\{  1,\dots,n\right\} \\
P_{n+1}\left(t\right) & := & Prob\left(I\left(.,Y_{t},t\right)=0\right)
\end{eqnarray*}
is generated by a CPH model with the transition matrix
given as below:

\[
A=\begin{Bmatrix}-c-\lambda_{1} & \lambda_{1} & 0 & \dots & 0 & c\\
0 & -c-\lambda_{2} & \lambda_{2} & \dots & 0 & c\\
0 & 0 & \ddots & \ddots & \vdots & \vdots\\
\vdots & \ddots & \ddots & \ddots & \lambda_{n-1} & c\\
0 & 0 & \dots & 0 & -c & c\\
0 & 0 & \dots & 0 & 0 & 0
\end{Bmatrix}
\]
\end{theorem}

The proof of Theorem \ref{connection to CPH} is presented in \ref{proof main}. It turns out that for fixed $\tilde{q}$,  Condition \ref{condition 2.6.9} gives an advection equation of $\tilde{\rho}$, from which the desired functional form of $\tilde{\rho}$ can be solved under some boundary condition. In fact, using the characteristic method \cite{evans2010partial}, we can express the function $\tilde{\rho}$ as below: 

\begin{equation}\label{rho tilde form}
\tilde{\rho}\left(y,t\right)=
\tilde{\rho}_{b}
\left(\tilde{g}\left(y,t,s^{\ast}\left(
y,t\right)\right),s^{\ast}
\left(y,t\right)\right)\cdot exp\left(
-c\cdot t\right)
\end{equation} 
where we assume that the value of function $\tilde{\rho}$ on a given boundary curve $b$ is known and denoted as $\tilde{\rho}_{b}(.,.)$, and $s^{\ast}$ is the first time when the solution trajectory ($\tilde{g}$) of the Initial Value Problem (IVP) \ref{ode2} (start from $\left(y,t\right)$) touches the boundary curve $b$.

The next corollary is a direct result of Theorem \ref{connection to CPH}. It extends the construction in Theorem \ref{connection to CPH} to a more general situation where the transition probability from different transient states to the absorbing state does not have to be identical. Therefore, it always possible to achieve an arbitrary CPH model from a RGRST model satisfying a generalized version of condition \ref{condition 2.6.9}.

\begin{corollary}\label{algorithm 1}
The smooth requirement on the function $\tilde{\rho}$ in Theorem \ref{connection to CPH} can be replaced by the following weaker condition:

($\mathbf{\ast}$) \textbf{The function $\tilde{\rho}$ is a continuous and almost everywhere differentiable function with respect to the standard Lebesgue measure on $\left[0,\infty\right)^2$}.

Under the condition ($\mathbf{\ast}$), for an arbitrary given CPH model represented by the following transition matrix:
\begin{equation}\label{CPH transition matrix}
A=\begin{Bmatrix}-c_1-\lambda_{1} & \lambda_{1} & 0 & \dots & 0 & c_1\\
0 & -c_2-\lambda_{2} & \lambda_{2} & \dots & 0 & c_2\\
0 & 0 & \ddots & \ddots & \vdots & \vdots\\
\vdots & \ddots & \ddots & \ddots & \lambda_{n-1} & c_{n-1}\\
0 & 0 & \dots & 0 & -c_n & c_n\\
0 & 0 & \dots & 0 & 0 & 0
\end{Bmatrix}
\end{equation}
and the initial probability mass vector $\alpha:=\left(\alpha_1,\dots,\alpha_n\right)$, there always exists a RGRST model together with a set of partition curves $\left\{C_0\equiv 0<C_1<\dots<C_n\equiv\infty\right\}$ such that the mapping: \[\left\{(y,t):\,C_{i-1}(t)\leq y\leq C_{i}(t)\right\}\mapsto Phase_i,\,i\in \{1,\dots,n\}\] converts the RGRST model to the given CPH model. 

Moreover, the desired RGRST model and the partition curves can be inductively constructed through Algorithm \ref{construct rho}.

\begin{algorithm}
\caption{Construct\_$\tilde{\rho}$}\label{construct rho}
\begin{algorithmic}
\Require $\lambda=\left(\lambda_1,\dots,\lambda_{n-1}\right),\,c=\left(c_1,\dots,c_n\right),\,\alpha=\left(\alpha_1,\dots,\alpha_{n}\right);$
\State $\textbf{Set }C_0\equiv 0, b:=[0,\infty)\times\{0\}\cup \{0\}\times[0,\infty), \tilde{\rho}_b\equiv 1;$  
\For {$i=1$ to $n$}
   \If {$i<n$} 
	 \State $\textbf{Set }C_i(0) \textrm{ by Eq. \ref{initial} and } \alpha_i;$
	 \State $\textbf{Set } PDE_i \textrm{ subject to } \tilde{\rho}_i|_b=\tilde{\rho}_b \textrm{ by replacing } c \textrm{ in Eq. \ref{condition 2.6.9} with } c_i;$
     \State $\textbf{Set }\tilde{\rho}_i:=exp\left(solve\left(
PDE_i\right)\right);$ 
     \State $\textbf{Set } IVP_i \textrm{ by replacing } \tilde{\rho}\textrm{ in Eq. \ref{condition (2)} with } \tilde{\rho}_i;$
     \State $\textbf{Set }C_i:=solve(IVP_i);$
	\State $\textbf{ReSet }b:=\{0\}\times[C_i(0),\infty)\cup\{C_{i}(t):\,t\in [0,\infty)\};$
	 \State $\textbf{ReSet }\tilde{\rho}_b(y,t):=
	 \begin{cases}
1 & \left(y,t\right)\in\left\{ 0\right\} \times[C_{i}\left(0\right),\infty)\\
\tilde{\rho}_{i-1}\left(y,t\right) & \left(y,t\right)\in\left\{ \left(C_{i}\left(t\right),t\right):\, t\in[0,\infty)\right\} 
\end{cases};$
   \Else
     \State $\textbf{Set }PDE_n \textrm{ subject to } \tilde{\rho}_n|_b=\tilde{\rho}_b\textrm{ by replacing } c \textrm{ in Eq. \ref{condition 2.6.9} with } c_n;$
     \State $\textbf{Set }\tilde{\rho}_n:=exp\left(solve(PDE_n)
     \right);$ 
   \EndIf
\EndFor 
\State $\textbf{Set }C_n:\equiv \infty ;$
\State $\textbf{Set }\tilde{\rho}:= \sum_{i=1}^{n}\mathbf{\mathbf{1}}_{C_{i-1}(t)\leq y \leq C_{i}(t)}\cdot \tilde{\rho}_i$ 
\State \Return $\tilde{\rho}$    
\end{algorithmic}\nonumber
\end{algorithm}
\end{corollary}

Notice that given the partition curves $\left\{C_{0}\equiv0<C_{1}<\dots<C_{N-1}
,C_{N}\equiv\infty\right\}$, we can define the \enquote{price} of the $i$th stage for each $i\in\{1,\dots,n\} $ as the following expectation:

\begin{equation}\label{price}
\begin{aligned}
Price_{i}(t):=&E\left(Y_t-C_{i-1}(t)\mid Y\in\left[C_{i-1}(t),C_{i}(t)\right),I\left(.,Y_t,t\right)=1\right)\\
=&\frac{\int_{C_{i-1}(t)}^{C_i(t)}\left(y-C_{i-1}(t)\right)\cdot\tilde{p}\left(y,t\right)\cdot
\tilde{\rho}\left(y,t\right)dy}{\int_
{C_{i-1}(t)}^{C_i(t)}\tilde{p}\left(y,t\right)\cdot
\tilde{\rho}\left(y,t\right)dy}.
\end{aligned}
\end{equation}
Intuitively, the \enquote{price} of the $i$th stage (represented as the interval bounded by $C_{i-1}(t)$ and $C_i(t)$) is just the average extra money that has to be charged at time $t$ in order to keep a patient in that stage. 

%It is also good to know the logarithmic version of the price of each medical stage which is defined in a similar way to Equation \ref{price}:
%\begin{equation}\label{pricelog}
%\begin{aligned}
%log\_Price_{i}(t):=&E\left(\ln\left(Y_t\right)-\ln\left(C_{i-1}(t)\right)\mid Y\in\left[C_{i-1}(t),C_{i}(t)\right),I\left(.,Y_t,t\right)=1\right)\\
%=&\frac{\int_{C_{i-1}(t)}^{C_i(t)}\left(\ln(y)-\ln\left(C_{i-1}(t)\right)\right)\cdot\tilde{p}\left(y,t\right)\cdot
%\tilde{\rho}\left(y,t\right)dy}{\int_
%{C_{i-1}(t)}^{C_i(t)}\tilde{p}\left(y,t\right)\cdot
%\tilde{\rho}\left(y,t\right)dy}.
%\end{aligned}
%\end{equation}
%Equation \ref{pricelog} measures the proportion of the extra money to be charged comparing to the money that has been charged by the current time $t$ given that patients would stay in the stage $i$ longer than time $t$. Therefore, the value $log\_Price$ captures both of the patient's willingness to pay and the effect of charge accumulation on that willingness.

\subsection{A Two-Stage Algorithm}
Corollary \ref{algorithm 1} implies a two-stage algorithm that use the real hospital charge and LOS data as input to estimate the underlying CPH model and the RGRST model from which the CPH model is derived (for short, we will call the pair of CPH model and RGRST model as the CPH-RGRST model).
%the following regression equations:
%\begin{equation}\label{structural equation}
%\begin{aligned}
%\ln Charge =& \theta_{y,0}+\theta_{y}\cdot X + \epsilon_{Y}\\
%\ln LOS = & \beta_{t,0}+\beta_{t}\cdot X' +\epsilon_{LOS},
%\end{aligned}
%\end{equation}
%where we assume that the exponential of the residual term, $e^{\epsilon_{LOS}}$, induces a CPH distribution, and the joint $\left(e^{\epsilon_{Y}},e^{\epsilon_{LOS}}\right)$ 
%has its PDF derived from a RGRST model constructed by Algorithm \ref{construct rho} from the given CPH distribution. 
A pseudo-code of the two-stage algorithm is given as in Algorithm \ref{twostage}. 
\begin{algorithm}
\caption{Two-Stage Full Maximum Likelihood (FML) Estimation}\label{twostage}
\begin{algorithmic}
\State $\textbf{STAGE\_1:}$
\Require $\textrm{Sample LOS Data } \left(t_1,\dots,t_K\right)$
\State $\textbf{Solve }max:\sum_{k=1}^{K} ln\, p_{CPH}\left(t_k;\lambda,c,\alpha\right);$%,\beta\right);$
\State \Return $\hat{\lambda},\,\hat{c},\,\hat{\alpha}.$%,\,\hat{\beta}.$
\end{algorithmic}
\begin{algorithmic}
\State $\textbf{STAGE\_2:}$ 
\Require $\textrm{Sample Charge-LOS Data } \left(\left(y_1,t_1\right),\dots,\left(y_K,t_K\right)\right)$ 
\State $\textbf{Apply } \textbf{STAGE\_1} \Longrightarrow \hat{\lambda},\,\hat{c},\,\hat{\alpha};$%,\,\hat{\beta};$
\State $\textbf{Set } \tilde{\rho}_{\hat{\lambda},\,\hat{c},\,\hat{\alpha}}:= \textbf{Construct\_}\tilde{\rho}\left(
\hat{\lambda},\,\hat{c},\,\hat{\alpha}\right);$
\State $\textbf{Set }\textrm{Joint PDF of Charge and LOS by Eq. \ref{joint density 2nd expression} and }\tilde{\rho}_{\hat{\lambda},\,\hat{c},\,\hat{\alpha}};$  
\State $\textbf{Solve } max: \sum_{k=1}^{K} ln\, PDF\left(y_k,t_k;\,params\right);$%,\theta\right);$
\State \Return $\hat{\lambda},\,\hat{c},\,\hat{\alpha},\,\widehat{params}.$%\hat{\beta},\,\hat{\theta},\,\widehat{params}.$
\end{algorithmic}
\end{algorithm}

In the first-stage estimation, we apply the FML method and the marginal LOS data to estimate the transition matrix and the initial probability mass that determines the marginal CPH distribution of LOS. The resulting estimators are denoted as %$\hat{\beta}:=\left(\hat{\beta}_{t,0},\hat{\beta}_t\right)$, 
$\hat{\lambda}:=\left(\hat{\lambda}_1,\dots,\hat{\lambda}_{N-1}\right)$, $\hat{c}:=\left(\hat{c}_1,\dots,\hat{c}_{N}\right)$ and $\hat{\alpha}:=\left(\hat{\alpha}_1,\dots,\hat{\alpha}_{N}\right)$.
 
In the second-stage estimation, we apply the Algorithm \ref{construct rho} to construct the function $\tilde{\rho}$ from the estimators obtained in the first stage and construct the joint PDF of charge and LOS by the formula \ref{joint density 2nd expression}. With the joint PDF, we construct the likelihood function and apply FML to estimate the remaining parameters, which are 
%
%Step 1: Use the estimator $\hat{\alpha}$ to construct the initial points of the partition curves;
%
%Step 2: Iteratively construct and solve (numerically) the ordinary differential equations for partition curves from the estimators $\hat{\lambda}$, $\hat{c}$ and $\hat{\alpha}$ as instructed in Equation \ref{condition (2)};
%
%Step 3: Using results in Step 1 and Step 2 to construct the empirical joint PDF and likelihood function in the way presented in \cite[]. Apply FML to estimate 
%the regression coefficients, $\hat{\theta}=\left(\hat{\theta}_{y,0},\hat{\theta}_{y}\right)$, and those parameters 
used to characterize the function $\tilde{q}$ and the initial density $p(.,0)$ (denote as $\widehat{params}$). 

The use of FML guarantees that all estimators obtained from the two-stage algorithm are consistent and asymptotically normal-distributed.

\section{Fitting Setup and Report}
\subsection{Parametrize $\tilde{q}$ and $p\left(.,0\right)$}
To apply the two-stage algorithm, we have to parametrize the initial density function $p\left(.,0\right)$ as well as the function $\tilde{q}$.

As suggested in \cite{xqzhang2017}, the functional form of $\tilde{q}$ is irrelevant with the distribution to be fitted, we can choose a simple form for the purpose of convenience. In this paper, we will consider the following parametric form:
\begin{equation}\label{q_tilde form}
\tilde{q}\left(y,t\right)= y.
\end{equation}
Consequently, the time-dependent PDF induced by $\{G_t\}$ can be represented as below:
\begin{equation}\label{p_tilde}
\tilde{p}\left(y,t\right) = p\left(y\cdot e^{-t},0\right)\cdot e^{-t}.
\end{equation}

Following \cite{gardiner2002longitudinal}, we assume the initial density function, $p\left(.,0\right)$, comes from a log-normal distribution, i.e.

\begin{equation}\label{log-normal density}
p\left(y,0\right)=\frac{1}{\sqrt{2\pi}\sigma \cdot y}\cdot exp\left(-\frac{\left(y-\mu
\right)^2}{2\sigma^2}\right)
\end{equation} 
It turns out that the parametrization given in Equation \ref{q_tilde form}, \ref{log-normal density} and \ref{rho tilde form} is nice in the sense of the three principles proposed in \cite{xqzhang2017}.

Combining Equation \ref{rho tilde form},  \ref{p_tilde}, the log-normal initial density as well as the joint PDF \ref{joint density 2nd expression}, we have the following parametric expression of the joint PDF of charge and LOS in our setting:
\begin{equation}\label{joint_d}
\Scale[0.9]{
P\left(y,t\right)=\frac{1}{\sqrt{2\pi}\sigma y}\cdot exp\left(-\frac{\left(ln y-t-\mu\right)^{2}}{2\sigma^{2}}\right)\cdot\sum_{i=1}^{n}
\mathbf{1}_{C_{i-1}
\left(t;\alpha,\lambda,c\right)\leq y<C_{i}\left(t;\alpha,\lambda,c\right)}\cdot\tilde{\rho}_{i}
\left(y,t;\alpha,\lambda,c\right)}
\end{equation}

Given a sample (size = $K$) of charge and LOS data, 
%$\left((y_1,los_1),\dots,(y_K,los_K)\right)$, 
$\left((y_1,t_1),\dots,(y_K,t_K)\right)$, 
the log-likelihood function and the maximization problem that has to be solved for in the second-stage estimation can be expressed as below:
%\begin{equation}\label{fml2}
%\begin{gathered}
%\Scale[0.8]{
%\max L^2\left(\mu,\sigma,\theta\right):=\sum_{k=1}^{K}\ln
%\left(
%\begin{aligned}
%&\frac{e^{\theta_{y,0}+\theta{y}x_k}}{\sqrt{2\pi}\sigma y_k}\cdot\exp\left(-\frac{\left(\ln y_k-\theta_{y,0}-\theta_{y}x_k-t_k-\mu\right)^{2}}{2\sigma^{2}}\right)\times\\
%&\sum_{i=1}^{n}\mathbf{1}_{C_{i-1}
%\left(t_k;\hat{\alpha},\hat{\lambda},\hat{c}\right)\leq y_k\cdot e^{-\theta_{y,0}-\theta_{y}x_k}<C_{i}
%\left(t_k;\hat{\alpha},\hat{\lambda},\hat{c}\right)}\cdot\tilde{\rho}_{i}
%\left(y_k\cdot e^{-\theta{y,0}-\theta_{y}x_k},t_k;\hat{\alpha},\hat{\lambda},\hat{c}\right)
%\end{aligned}
%\right) }\\
%\begin{aligned}
%s.t.\,\,\,&\sigma>0\\
%&t_k=los_k\cdot e^{-\hat{\beta}_{t,0}-\hat{\beta}_{t}x_k},\,k=1,\dots,K
%\end{aligned}
%\end{gathered}
%\end{equation}
\begin{equation}\label{fml2}
\begin{gathered}
\Scale[0.8]{
max\,\, L^2\left(\mu,\sigma\right):=\sum_{k=1}^{K} ln
\left(
\begin{aligned}
&\frac{1}{\sqrt{2\pi}\sigma y_k}\cdot exp\left(-\frac{\left(ln y_k-t_k-\mu\right)^{2}}{2\sigma^{2}}\right)\times\\
&\sum_{i=1}^{n}\mathbf{1}_{C_{i-1}
\left(t_k;\hat{\alpha},\hat{\lambda},\hat{c}\right)\leq y_k<C_{i}
\left(t_k;\hat{\alpha},\hat{\lambda},\hat{c}\right)}\cdot\tilde{\rho}_{i}
\left(y_k,t_k;\hat{\alpha},\hat{\lambda},\hat{c}\right)
\end{aligned}
\right) }\\
s.t.\,\,\,\sigma>0
\end{gathered}
\end{equation}
where the value of estimators, $\hat{\alpha}$, $\hat{\lambda}$ and $\hat{c}$, is fixed and comes from the first-stage estimation. The first-stage log-likelihood function and the associated maximization problem can be expressed as:

%\begin{equation}\label{fml1}
%\begin{gathered}
%\max L^1\left(\alpha,\lambda,c,\beta\right):=\sum_{k=1}^{K}\ln \alpha\cdot e^{A(\lambda,c)\cdot t_k}\cdot A(\lambda,c)\cdot 1_{n}\\
%\begin{aligned}
%s.t.\,\,\,\,&\lambda_i>0,\,\\
%& 0<c_i<\lambda_i,\,i=1,\dots,n-1;\\
%&c_n>0,\\
%&t_k=los_k\cdot e^{-\beta_{t,0}-\beta_{t}x_k},\,k=1,\dots,K.
%\end{aligned}
%\end{gathered}
%\end{equation}
\begin{equation}\label{fml1}
\begin{gathered}
max\,\, L^1\left(\alpha,\lambda,c\right):=\sum_{k=1}^{K} ln \left( \alpha\cdot e^{A(\lambda,c)\cdot t_k}\cdot A(\lambda,c)\cdot 1_{n}\right)\\
\begin{aligned}
s.t.\,\,\,\,&\lambda_i>0,\,i=1,\dots,n-1;\\
& c_i>0,\,i=1,\dots,n;\\
&\sum_{i=1}^{n} \alpha_i=1.
\end{aligned}
\end{gathered}
\end{equation}
where $A(\lambda,c)$ is the $n-$dimensional transition matrix determined by $\lambda$ and $c$ as in Equation \ref{CPH transition matrix}. $1_n$ is the $n-$dimensional vector with all entries being 1. Following \cite{Tang2012thesis} and \cite{mcclean2005markov}, we will use a four-phase CPH model for the estimation in the first stage. Therefore, $n=4$ in \ref{fml1}.

\subsection{Fitting Results}
In this section, we will apply the two-stage algorithm to estimate 
%the regression coefficients $\vec{\theta}=\left(\hat{\theta}_{y,0},\hat{\theta}_{y}\right)$, $\vec{\beta}=\left(\hat{\beta}_{t,0},\hat{\beta}_{t}\right)$, and 
the parameter vectors $\left(\mu,\sigma\right)$, $\lambda$, $c$ and $\alpha$ where $\mu$ and $\sigma$ are the geometric mean and standard deviation of the log-normal initial distribution. \\

\begin{minipage}{\linewidth}
\begin{center} 
\captionof{table}{Estimated Dynamic Parameters}\label{table: dynamic para}
\centering
\resizebox{9.5cm}{1.4cm}{%
\begin{tabular}{ll}
\hline 
\hline
Dynamic Parameters & Values\tabularnewline
\hline 
($\mu$,$\sigma$) & (-0.5715, 0.7149)\tabularnewline
$\alpha$ & (0.99972043, 0.0000001, 0.0000001, 0.00027937)\tabularnewline
c & (-2.09, -9.05,-0.91,-0.14)\tabularnewline
$\lambda$ & ($\approx$0.00,6.45, 0.83)\tabularnewline
\hline 
\hline 
 &  \tabularnewline
\end{tabular}%
}
\end{center}
\end{minipage}
The data we used for estimation is a sample (size = 5000) drawn from the New York State's Statewide Planning and Research Cooperative System 2013 (SPARCS 2013). The detailed property of the data has been discussed in \cite{xqzhang2017}. 
The in-sample fitting results are plotted in Figure \ref{fig: in-sample}. 
%The estimated value and the associated Pvalue of regression coefficients are reported in Table \ref{table: coefficient}. 
Estimated values of parameters are reported in Table \ref{table: dynamic para}. The goodness of fitting is measured by Pearson's $\chi^2$, the value of the $\chi^2$ statistics and the associated P-values are (0.0171,1.0) for the marginal charge, (6.8764,0.7371) for the marginal LOS and (0.1911,1.0) for the joint. From the goodness-of-fit plot \ref{fig: in-sample} and the $\chi^2$ test, we can conclude that the fitting generated by the four-phase CPH-RGRST model to the sampled data is very good.\\

\begin{minipage}\linewidth
\begin{center}
%\centering
\captionof{figure}{In-Sample Marginal/Joint Fitting of Log-charge and LOS by RGRST Model}\label{fig: in-sample}
\includegraphics[width=12cm,height=10cm]{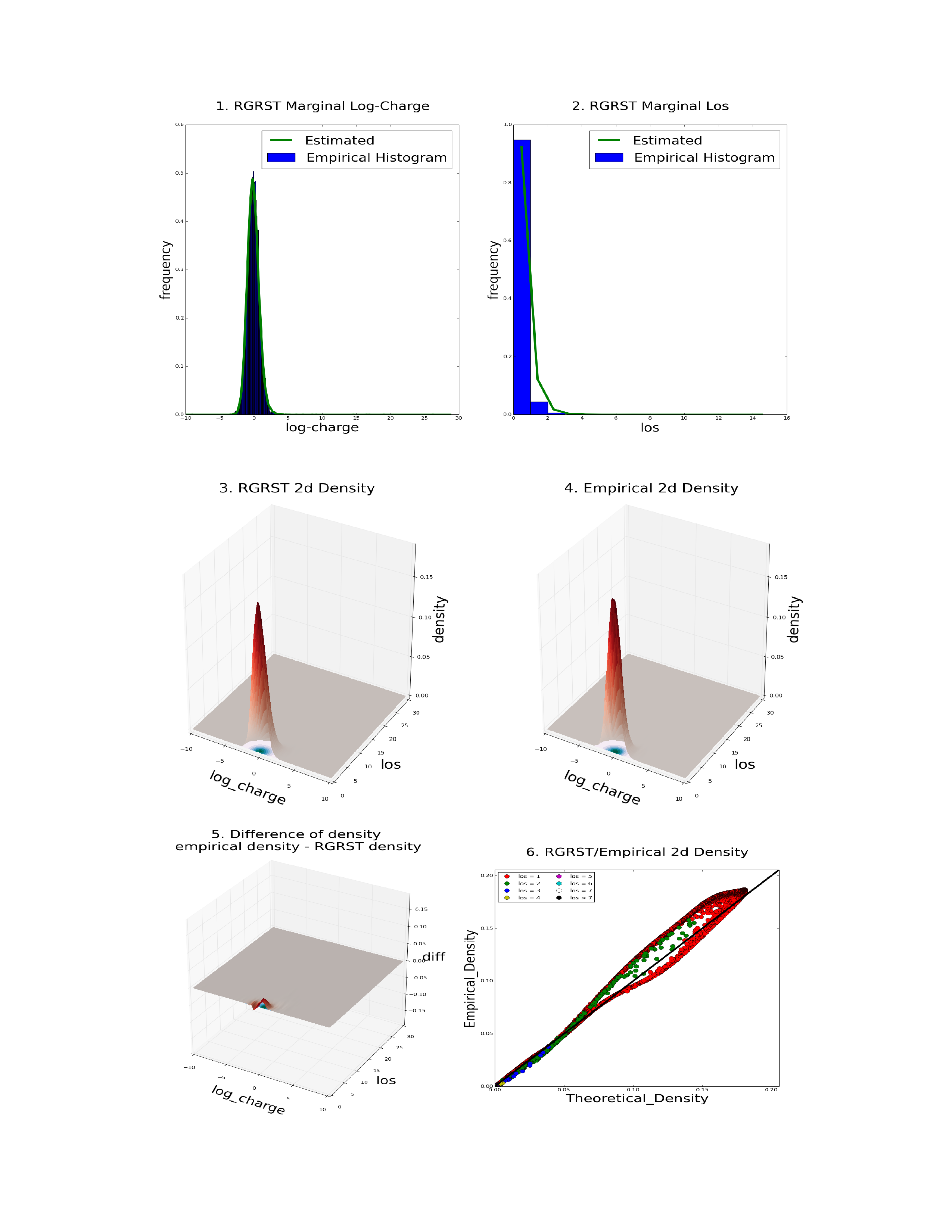}
\begin{minipage}{1\textwidth}
{\footnotesize Plot 1 and 2 are the marginal fitted RGRST distribution v.s. empirical histogram for log-charge and LOS. Plot 3 is the joint density of log-charge and LOS derived from the fitted RGRST model. Plot 4 is the empirical joint density obtained from Gaussian kernel density estimation (KDE) with kernel width 0.15 for log-charge and 1 for LOS. Plot 5 is obtained from subtracting Plot 3 from Plot 4. Plot 6 is the KDE density versus the RGRST density evaluated at all 5000 samples.\par}
\end{minipage}
\end{center}
\end{minipage}\\

To avoid \enquote{over-fitting}, we also apply Pearson's $\chi^2$ test to the out-sample fitting result where the out-sample is chosen to be the complement dataset of the 5000 sample records within SPARCS 2013 which contains millions of records. The $\chi^2$ statistics and P-values for out-sample fitting are (0.0112,1.0) for the marginal charge, (6.8068,0.7435) for the marginal LOS and (0.2457,1.0) for the joint (out-sample fitting results are plotted in Figure \ref{fig: out-sample}). Comparing Pearson's $\chi^2$ computed for the in-sample and out-sample fitting, we find that there is no significant difference between them, which indicates no \enquote{over-fitting}.\\

\begin{minipage}\linewidth
\begin{center}
%\centering
\captionof{figure}{Out-Sample Marginal/Joint Fitting of Log-charge and LOS by RGRST Model}\label{fig: out-sample}
\includegraphics[width=12cm,height=10cm]{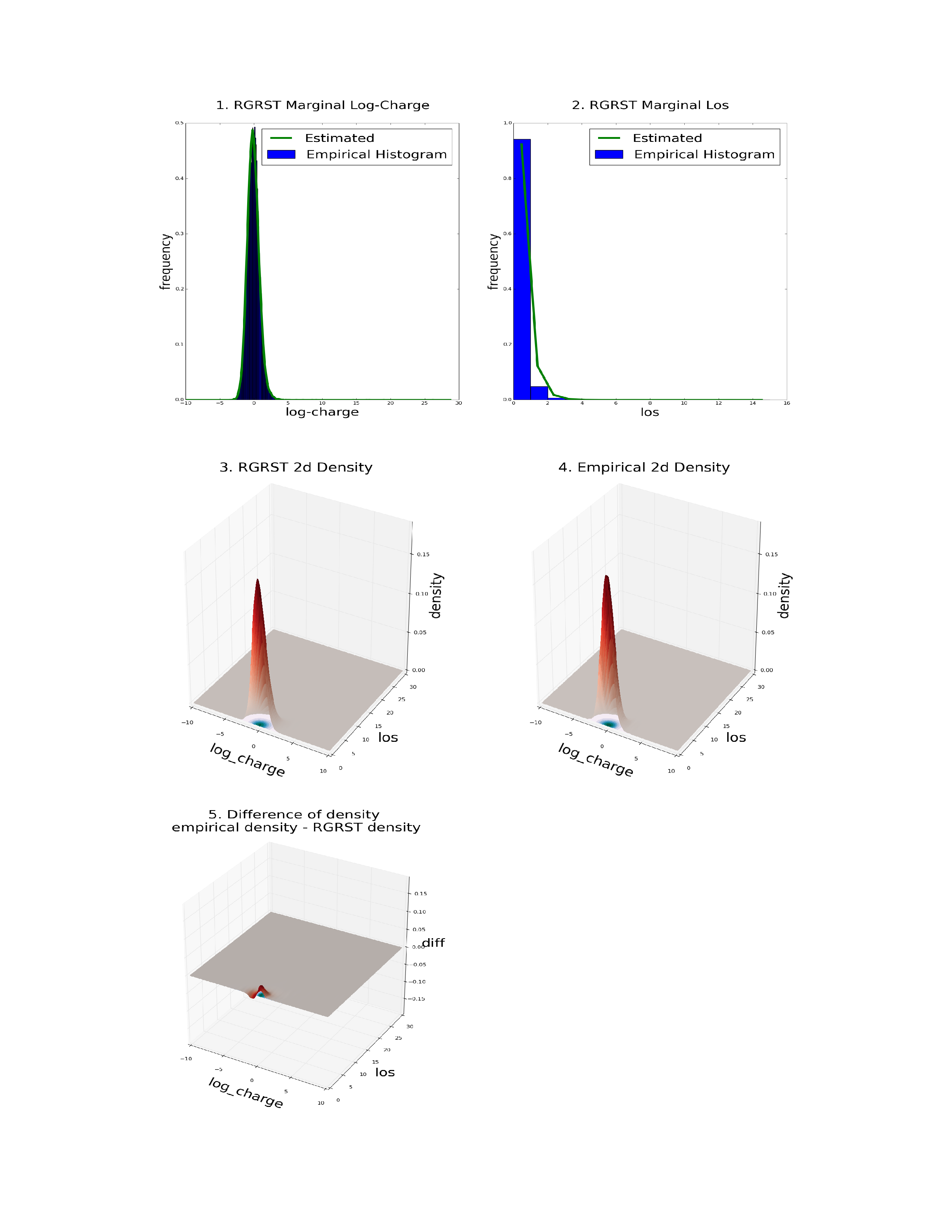}
\begin{minipage}{1\textwidth}
{\footnotesize Plot 1 and 2 are the marginal fitted RGRST distribution v.s. empirical histogram for log-charge and LOS. Plot 3 is the joint density of log-charge and LOS derived from the fitted RGRST model. Plot 4 is the empirical joint density obtained from Gaussian kernel density estimation (KDE) with kernel width 0.15 for log-charge and 1 for LOS. Plot 5 is obtained from subtracting Plot 1 from Plot 2.\par}
\end{minipage}
\end{center}
\end{minipage}

\subsection{Price}
As discussed in the end of the section 2.1, the CPH-RGRST model enables us to evaluate the price (as defined in Equation \ref{price}) of each medical stage. Using the estimation results provided in the previous section, we can even estimate this price from the charge and LOS data given in SPARCS 2013.

In Figure \ref{fig:price}, we plot the estimated price curves of the four phases in the fitted CPH model. From Figure \ref{fig:price}, the price of all stages are increasing over time which is consistent with the fact that the longer hospital stay would consume more medical resource and therefore induce higher charge. 

In addition, comparing to other stages, the stage 1 and 4 are always quite expensive. One possible explanation is that the stage 1 is the first stage after admission and during this stage a lot of physical examinations, experimental treatments and procedures will be applied, all of which are quite expansive and lift up the total cost for that stage. In contrast, the stage 4 is the last potential stage that a patient could experience before get discharged, meanwhile patients who weren't discharged before reaching the stage 4 are more likely to suffer from very severe illness and therefore need more care before discharge. Consequently, many good-quality but expensive medicines and procedures will be applied for the recovery purpose, which makes the stage 4 expensive. 

It is also notable that the price of the 3rd stage increases drastically. Especially as the time exceeds 15 days, the stage 3 becomes even more expensive than the stage 1 and 4. One possible reason to that phenomenon is that patients with relatively long stay and reaching the 3rd stage in a late time are more likely to experience a major procedure in the stage 3, like a serious surgery. The major procedure itself is expensive and the high price of the stage 3 is just a reflect of its cost.\\

%Figure \ref{fig: price} shows the time-dependent price of four medical stages identified with the four phases in the estimated CPH model and how these price curves could vary along with the increasing of APR Severity of Illness Code (the code takes integer value within $\{1,2,3,4\}$ and the greater value of the code indicates the more complex illness condition and the more medical resources to be used).

\begin{minipage}\linewidth
\begin{center}
%\centering
\captionof{figure}{Log-Price by Phase}\label{fig:price}      
\includegraphics[width=10cm,height=6cm]{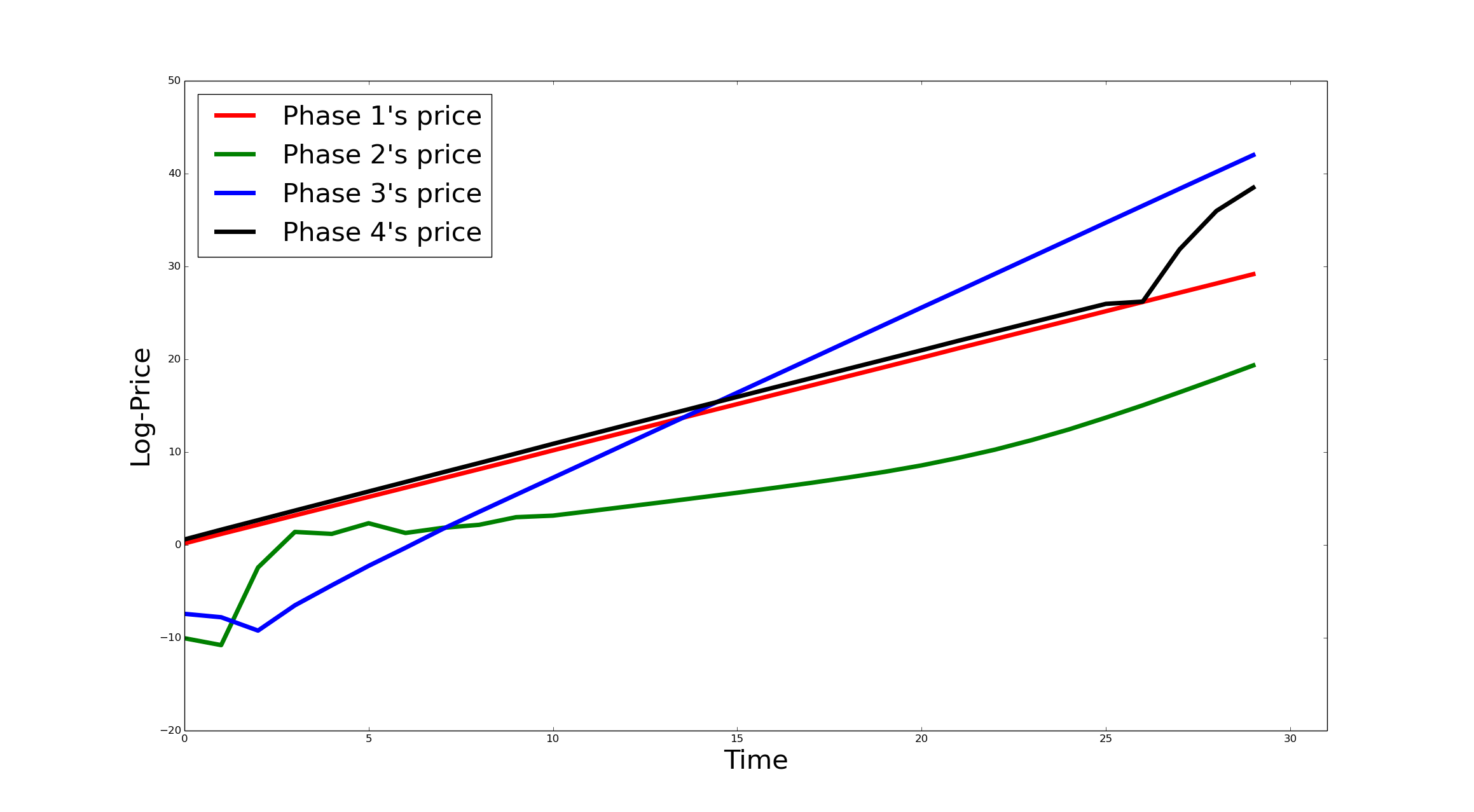}

\end{center}
\end{minipage}

\section{Discussion}
We have described a methodology whereby the widely used CPH models and RGRST models can be combined together and the price of each phase in the CPH model can be defined. We use the charge and LOS data sampled from SPARCS 2013 to estimate the price curves of each medical stage in a four-phase CPH-RGRST model. There are a couple of interesting directions that can be done in future researches:

First, the estimated price curves can be used together with the time-dependent probability of patients staying in each medical stage \cite{mcclean2005markov}, which can be derived directly from the first-stage estimation in Algorithm \ref{twostage}. The availability of the probability curves of all stages and their associated price curves facilitates monitoring the cost evolution during hospital stay and makes it convenient to control the cost variation across different stages. Therefore, the CPH-RGRST models provide an efficient cost-analysis tool for the treatment dynamics in hospital.

It is also good to include the effect of covariates on the price curves that can be easily done by setting up regression equations that link the covariates with the parameters involved in the CPH-RGRST models as did in \cite{Tang2012thesis,
mcclean2005markov,
gardiner2002longitudinal}. The inclusion of the regression analysis helps identify different patient groups by patient's personal characteristics (like age, gender, diagnosis-related features) and facilitate the application of our cost analysis to specific patient groups. In fact, for different patient groups, the price curves may display significantly different features, which are important to understand many welfare-related problems of special patient groups.

\begin{appendices}
\section{Proof of Theorem \ref{connection to CPH}}\label{proof main}
\begin{proof}
The main idea of the proof is to construct the partition $\mathcal{P}$
by induction. Assume, firstly, the partition $\mathcal{P}$ is formed
by n curves in $[0,\infty)$ (denoted as $\left(t,C_{i}\left(t\right)\right)$
for $i\in\left\{ 0,1,\dots,n-1\right\} $) satisfying increasing condition
as below: 
\begin{equation}
\label{increasing condition}
\begin{gathered}
C_{i+1}\left(t\right) > C_{i}\left(t\right)\geq0\\
C_{0}\left(t\right) \equiv 0
\end{gathered}
\end{equation}
 in such a way that $\mathcal{P}_{i}:=\left\{ \left(y,t\right):C_{i-1}\left(t\right)\leq y<C_{i}\left(t\right)\right\} $
for $i\in\left\{ 1,\dots,n\right\} $ (For simplicity of notation,
we assume $C_{n}\left(t\right):\equiv\infty$.). So, to prove the
theorem, it suffices to find out a family of increasing curves $\left\{ C_{i}:i\in\left\{ 0,1,\dots,n-1\right\} \right\} $
with the induced transition matrix is as stated in the theorem.

Firstly, notice that by condition \ref{condition 2.6.9}, we have for $i\in\left\{ 0,1,\dots,n-1\right\} $
and any family of n curves $\left\{ C_{i}:i\in\left\{ 0,1,\dots,n-1\right\} \right\} $
satisfying increasing condition \ref{increasing condition}, the following holds:

\begin{equation}
\begin{gathered}
\lim_{\delta\downarrow0}
\frac{Prob\left(I\left(.,Y_{t+\delta},t+\delta\right)=0|Y_{t}\in[C_{i}\left(t\right),C_{i+1}\left(t\right)),I\left(.,Y_{t},t\right)=1\right)}{\delta}\\
=\frac{\int_{C_{i}\left(t\right)}^{C_{i+1}\left(t\right)}-\left(\frac{\partial\tilde{\rho}}{\partial y}\cdot\tilde{q}+\frac{\partial\tilde{\rho}}{\partial t}\right)\left(y,t\right)\cdot\tilde{p}\left(y,t\right)dy}{\int_{C_{i}\left(t\right)}^{C_{i+1}\left(t\right)}\tilde{\rho}
\left(y,t\right)\cdot\tilde{p}
\left(y,t\right)dy} \equiv c.
\end{gathered}
\end{equation}

Moreover, because

\begin{equation}
Prob\left(Y_{t}\geq C_{n}\left(t\right)=+\infty,I\left(.,Y_{t},t\right)=1\right)
=0,
\end{equation}

we can conclude that no matter the choice of $\left\{ C_{i}:i\in\left\{ 0,1,\dots,n-1\right\} \right\} $,
the transition probability matrix $A$ always has its column n+1 of
the following form:

\[
A_{n+1}=\begin{Bmatrix}c\\
\vdots\\
c\\
0
\end{Bmatrix}
\]

On the other hand, the increasing condition and the
Non-Increasing property \ref{non-increasing property} of RGRST processes, we have for all $i,j\in\left\{ 0,1,\dots,n-1\right\} $
with $j\not=i$ or $i+1$ and small enough $\delta$ 

\[
Prob\left(Y_{t+\delta}\in[C_{j}\left(t\right),C_{j+1}\left(t\right))
\mid Y_{t}\in[C_{i}\left(t\right),C_{i+1}\left(t\right)),I\left(.,Y_{t},t\right)=1\right)=0
\]

and 

\[
Prob\left(Y_{t+\delta}\in[C_{j}
\left(t\right),C_{j+1}\left(t\right)),
I\left(.,Y_{t+\delta},t+\delta\right)=1
\mid I\left(.,Y_{t},t\right)=0\right)=0.
\]

That is, the induced transition matrix $A$ satisfies 

\[
A_{i,j}=0,\, j\not\in\left\{i,i+1\right\}\text{ and } i,j\in\left\{ 1,\dots,n\right\} 
\]

and 

\[
A_{n+1,j}=0\textrm{, }j\in\left\{ 1,\dots,n\right\} 
\]

So, to prove the theorem, it suffices to construct a family of increasing curves
$\left\{ C_{i}:i\in\left\{ 0,1,\dots,n-1\right\} \right\} $ guaranteeing
that 

\begin{eqnarray*}
A_{i,i+1} & = & \lambda_{i}\textrm{, }i\in\left\{ 1,\dots,n-1\right\} 
\end{eqnarray*}

and 

\[
P_{i}\left(0\right)=\alpha_{i}\textrm{, }i\in\left\{ 1,\dots,n-1\right\} 
\]

However, these two conditions are equivalent to finding:

(1) A sequence $\left\{ c_{i}:i\in\left\{ 0,1,\dots,n\right\} \right\} $
with $c_{0}=0$ and $c_{n}=\infty$ such that 

\begin{equation}\label{initial}
\alpha_{i}=\int_{c_{i-1}}^{c_{i}}p\left(y,0\right)dy\textrm{, }i\in\left\{ 1,\dots,n-1\right\}; 
\end{equation}

(2) a sequence of solutions $\left\{ y_{i}\left(t\right):i\in\left\{ 0,1,\dots,n-1\right\} \right\} $
with $y_{0}\equiv0$ to initial value problems (IVP) for $i\in\left\{ 1,\dots,n-1\right\} $

\begin{equation}
\label{condition (2)}
\begin{aligned}
\frac{dy_{i}}{dt} & =  \tilde{q}\left(y_{i},t\right)-\lambda_{i}\cdot\frac{\int_{y_{i-1}}^{y_{i}}\tilde{p}\left(x,t\right)\cdot\tilde{\rho}\left(x,t\right)dx}{\tilde{p}\left(y_{i},t\right)\cdot\tilde{\rho}\left(y_{i},t\right)}\\
y_{i}\left(0\right) & = c_{i}
\end{aligned}
\end{equation}

Condition \ref{condition (2)} arises from the following equality 

\begin{equation}
\Scale[0.8]{
\begin{aligned}
&\lim_{\delta\downarrow0}\frac{Prob\left(Y_{t+\delta}\in[C_{i}\left(t+\delta\right),C_{i+1}\left(t+\delta\right)),I\left(.,Y_{t+\delta},t+\delta\right)=1|Y_{t}\in[C_{i-1}\left(t\right),C_{i}\left(t\right)),I\left(.,Y_{t},t\right)=1\right)}{\delta} \\
=&\lim_{\delta\downarrow0}\frac{Prob\left(Y_{t+\delta}\in[C_{i}\left(t+\delta\right),C_{i+1}\left(t+\delta\right)),I\left(.,Y_{t+\delta},t+\delta\right)=1,Y_{t}\in[C_{i-1}\left(t\right),C_{i}\left(t\right)),I\left(.,Y_{t},t\right)=1\right)}{Prob\left(Y_{t}\in[C_{i-1}\left(t\right),C_{i}\left(t\right)),I\left(.,Y_{t},t\right)=1\right)} \\
=&\lim_{\delta\downarrow0}\frac{\int_{C_{i}\left(t+\delta\right)}^{\tilde{g}^{-1}\left(C_{i}\left(t\right),t,t+\delta\right)}\tilde{p}\left(y,t+\delta\right)\cdot\tilde{\rho}\left(y,t+\delta\right)dy}{\int_{C_{i-1}\left(t\right)}^{C_{i}\left(t\right)}\tilde{p}\left(y,t\right)\cdot\tilde{\rho}\left(y,t\right)dy}  \\
=&\frac{\tilde{p}\left(C_{i}\left(t\right),t\right)\cdot\tilde{\rho}\left(C_{i}\left(t\right),t\right)\cdot\left(\lim_{\delta\downarrow0}\frac{\tilde{g}^{-1}\left(C_{i}\left(t\right),t,t+\delta\right)-\tilde{g}^{-1}\left(C_{i}\left(t\right),t,t\right)}{\delta}-\frac{dC_{i}\left(t\right)}{dt}\right)}{\int_{C_{i-1}\left(t\right)}^{C_{i}\left(t\right)}\tilde{p}\left(y,t\right)\cdot\tilde{\rho}\left(y,t\right)dy}  \\
=&\frac{\tilde{p}\left(C_{i}\left(t\right),t\right)\cdot\tilde{\rho}\left(C_{i}\left(t\right),t\right)\cdot\left(\tilde{q}\left(C_{i}\left(t\right),t\right)-\frac{dC_{i}\left(t\right)}{dt}\right)}{\int_{C_{i-1}\left(t\right)}^{C_{i}\left(t\right)}\tilde{p}\left(y,t\right)\cdot\tilde{\rho}\left(y,t\right)dy} = \lambda_{i},
\end{aligned}}\nonumber
\end{equation}
where the time-dependent density function $\tilde{p}$ of the potential charge process $\{G_t\}$ has the form of $\tilde{p}=p\left(\tilde{g}\left(y,t
,t\right),0\right)\cdot \frac{\partial\tilde{g}\left(y,t,t\right)}{\partial y}$. The involved function $\tilde{g}$, viewed as a family of functions in
variable $s$, solves the following family of IVPs:

\begin{equation}
\label{ode2}
\begin{aligned}
\frac{dy}{dt} =  -\tilde{q}\left(y,t-s\right)\\
\tilde{g}\left(y,t,0\right) =  y
\end{aligned}
\end{equation}
and $\tilde{g}^{-1}$ is defined as the inverse to $\tilde{g}$ such
that $\tilde{g}^{-1}\left(y,0,t\right):=\left\{ x:\tilde{g}\left(x,t,t\right)=y\right\}$.

It is easy to check condition (1) does always hold. Finally, thanks to the
existence and uniqueness theorem of solutions to IVPs, the solution curves $$\left\{ y_{i}\left(t\right):i\in\left\{ 0,1,\dots,n-1\right\} \right\} $$
do exist to the IVPs in \ref{condition (2)}. 
\end{proof}
\end{appendices} 
%%%%%%%%%%%%%%%%%%%%%%%%%%%%%%%%%%%%%%%%%%%%%%%%%%%%%%%%%%%%
%%%%%%%%%%%%%%%%%%%%%%%%%%%%%%%%%%%%%%%%%%%%%%%%%%%%%%%%%%%%%%%%%%%%%%%%%%%%%%%%%%%%%%%%%%%%%%%%%%%%%%%%%%%%%%%%%%%%%%%
%\renewcommand{\footnotesize}{\fontsize{8pt}{10pt}\selectfont}
%\bibliographystyle{unsrtnat}
%\bibliography{bib}

\end{document}